\documentstyle[aps]{revtex}
\input{epsf}
\def\dofig#1#2{\epsfysize=#1 \centerline{\epsfbox{#2}}}

\begin{document}
\title{Delay estimation from noisy time series}

\author{Toru Ohira}
\address{
Sony Computer Science Laboratory\\
3-14-13 Higashi-gotanda, Shinagawa,\\
 Tokyo 141, Japan
}

\author{Ryusuke Sawatari}
\address{
Department of Computer Science\\
Keio University\\
3-14-1 Hiyoshi, Yokohama 223, Japan
}

\date{\today}
\maketitle
\begin{abstract}
We propose here a method to estimate
a delay from a time series taking 
advantage of analysis
of random walks with delay.
 This method is
applicable to a time series coming out of a system
which is or can be approximated as a linear feedback
system with delay and noise. We successfully test the method
with a time series generated by discrete Langevin
equation with delay.
\end{abstract}

Estimation of delay from a noisy time series has attracted much attention.
Especially, when the time series is chaotic, estimation of delay
has a practical motivation: time--delayed
coordinates are typically used to estimate fractal dimensions and Lyapunov
exponents.
There are series of works considering the subject from 
this viewpoint \cite{takens,grassberger-wolf,bunner,sano-eckmann-tanaka}.
Another viewpoint is to consider that a noisy time series consists 
of underlying deterministic dynamics with past influence and
a noise term.
Some statisticians have taken this stand and devised
methods of analysis, for example, using the generalized
Langevin equation \cite{kubo,mori} and
fluctuation dissipation theorem\cite{okabe}.
In physiological fields, 
a more specialized case of the feedback delay associated
with control system has attracted great deal of interest. A series of
attempts has been made to estimate delay from physiological experimental 
data (see e.g., \cite{milton,ohira1,eurich}).

Against this background, we present here a method of estimating delay
from a time series which is or is approximately generated by
a noisy linear feedback system. We take advantage
of analysis of random walks whose transition 
probability depends on the walker's position in a fixed interval
past. Such random walks are termed delayed random walks and were proposed 
recently as a platform on which to study 
systems with both noise and delay \cite{ohira1,ohira2}.
We will describe each step of the method in a transparent manner for 
implementation into computer algorithms. The method is tested to
show its effectiveness on several test time series generated by
discrete Langevin equation with delay \cite{kuchler}. 

Let us first describe the delayed random walk, on whose
analysis we base our method for delay estimation.
We consider a random walk which takes a unit step in a unit time.
The delayed random walk we start with is an extension of a position 
dependent random walk whose step toward the origin is more likely
when no delay exists. 
Formally, it has the following definition:
\begin{eqnarray}
P(X_{t+1}&=&n; X_{t+1-\tau}=s)\nonumber\\
 &=& g(n-1,s-1)P(X_{t}=n-1;X_{t+1-\tau}=s;X_{t-\tau}=s-1)\nonumber\\ 
 &+& g(n-1,s+1)P(X_{t}=n-1;X_{t+1-\tau}=s;X_{t-\tau}=s+1)\nonumber\\
 &+& f(n+1,s-1)P(X_{t}=n+1;X_{t+1-\tau}=s;X_{t-\tau}=s-1)\nonumber\\
 &+& f(n+1,s+1)P(X_{t}=n+1;X_{t+1-\tau}=s;X_{t-\tau}=s+1),\\
f(x, y) &= & {1 \over 2}(1 + \alpha x + \beta y)\nonumber\\
g(x, y) &= & {1 \over 2}(1 - \alpha x - \beta y)
\label{drw}
\end{eqnarray}
where the position of 
the walker at time $t$ is $X_{t}$, and $P(X_{t_1}=u_1;X_{t_2}=u_2)$ 
is the joint probability
for the walker to be at $u_1$ and $u_2$ at time $t_1$ and $t_2$, respectively. 
$f(x, y)$ and $g(x, y)$ are transition probabilities to take a step to the negative 
and positive directions respectively. 
Hence the transition probability depends on both the current and the
$\tau$ steps past positions of the walker. We note that the above definition
is approximate: we are assuming that the probability for the walker
to be at positions which violate the condition $ 0 \leq f(x,y) \leq 1$ 
is negligible. This is an extension of the delayed random walk model
discussed in \cite{ohira2}.

Following the similar argument as \cite{ohira2}, we can derive a 
set of 
coupled equations which the stationary correlation function of the
model obeys:
\begin{eqnarray}
K(0) & = & (1-2\alpha)K(0) + 1 - 2\beta K(\tau)\nonumber\\
K(u) & = & (1- \alpha)K(u-1) - \beta K(\tau-(u-1)),\quad (1\leq u \leq \tau)\nonumber\\ 
K(u) & = & (1- \alpha)K(u-1) - \beta K((u-1)-\tau),\quad (u > \tau) 
\label{dynaeq}
\end{eqnarray}
We can solve this set of equations iteratively for $K(u)$ given
$\alpha$, $\beta$ and $\tau$; examples are shown in Figure 1.
We note that the oscillatory solutions are seen with sufficiently
large $\tau$, but the shapes of the curves are different 
for the cases of $\alpha > \beta$ and $\beta > \alpha$ \cite{sawatari}.

The delayed random walk model presented here can be considered 
a model of a linear delayed feedback system with noise
 in probability space.
This can be seen more transparently by considering its counterpart
in physical space, which is given as follows in discrete time with
white noise $\xi_{t}$ \cite{kuchler}.
\begin{equation}
X_{t+1} - X_{t} = - \alpha X_{t} - \beta X_{t-\tau} + \xi_{t}, \quad
\langle \xi_{t_1}\xi_{t_2}\rangle = \delta(t_1 - t_2).
\label{dle}
\end{equation}
If the system which generates a time series is or is approximated as
a noisy linear feedback system, we can use the above set of equations
to estimate the delay and other parameters. The basic idea is to 
``tune'' each parameter so that the correlation function from the
time series numerically matches the solution of 
equation $(\ref{dynaeq})$. We can derive several concrete
algorithms of different
approaches based on this basic idea. In the following we present
one such method which is simple with respect to both its concept and
its implementation. The concrete steps are as follows.

(1) As a prerequisite, we need to have a stationary noisy time series and
some physical assumption that it is or is approximately generated by 
a linear feedback
system with delay. (Some aspects of a time series such as 
whether it is chaotic or not
 can be checked
by already known methods \cite{grassberger-wolf}.)
 
(2)  Construct the auto--correlation function $C(u)$ from the time series. 
If it is oscillating with some $C(u)<0$, we can go to
the next step. (If not oscillating and $C(u)>0$, it is still possible
to ``tune'' parameters in principle. However, as other methods may 
be more appropriate\cite{okabe}, we do not consider this case here.)
An example is shown in Figure 2.

(3) From $\{ C(u) \}$, we generate a ``normalized set''. Decide on the
unit step size, and normalize the correlation function by the
following requirement derived from (\ref{dynaeq}).
\begin{equation}
K(0) - K(1) = {1 \over 2}
\end{equation}
Hence, we generate
\begin{equation}
K(u) = { C(u)\over {2(C(0) - C(1))} }
\end{equation}
We assume that with correctly estimated parameters, $K(u)$ generated this
way obeys (\ref{dynaeq}).

(4) Estimate delay $\tau_e$ around the ``first zero'' $\tau_i$ of the
correlation function; $\tau_i$ is the smallest number such that
$K(\tau_i)\approx 0$. 

(5) With estimated $\tau_e$, generate the following two sets of
ordered pair $L_1 = \{(y_1(u), z_1(u))\}$ and $L_2 = \{(y_2(u), z_2(u))\}$
 from $K(u)$
\begin{eqnarray}
y_1(u) &=& { K(u) \over K(|\tau_e - u|) }, \quad  
z_1(u)  =  { K(u+1) \over K(|\tau_e - u|) }\\ 
y_2(u) &=& { K(|\tau_e - u|) \over K(u) }, \quad  
z_2(u)  =  { K(u+1) \over K(u) }
\end{eqnarray}

(6) Plot $L_1$ and $L_2$. We use the following relation derived from
(\ref{dynaeq}):
\begin{equation}
z_1(u) = (1-\alpha) y_1(u) - \beta, \quad z_2(u) = -\beta y_2(u) - (1-\alpha)
\end{equation}
Thus, our assumption here is that if we have a correct estimate of
$\tau$ then both plots will be fitted with a linear function whose
slope and intercept will give us $\alpha$ and $\beta$.
Example of these plots are shown in Figure 3.

(7) Compute $\chi^2$ error for each plot, and define
\begin{equation}
E(\tau_e) = \chi^2_1 + \chi^2_2.  
\end{equation}
Our best estimate of $\tau$ is the one which minimizes
$E(\tau_e)$ near $\tau_i$. Corresponding $\alpha$ and $\beta$
is obtained as described in (6) (Figure 4).

We have tried this algorithm on several test time series data generated by 
equation (\ref{dle}) with various parameter ranges, and
sample
results are shown in Table 1. We found that choice of
the number of correlation function points used, $u_{max}$, 
occasionaly affect our results.
We heuristically chose $u_{max}$ at a value up to
which the graph of $C(u)$ is rather clear, typically about
two to four times $\tau_i$. The estimates are quite 
reasonable as shown here and typically better than
``first zero'' estimate $\tau_i$.

We have described a method of estimating parameters from
a time series produced by a noisy linear feedback system with
a single stable point using analysis of delayed random walks.
As mentioned, other algorithms based on the same
basic idea of ``tuning'' in to the correlation function
can be devised. A scheme of cross--checking estimated parameters from these 
different algorithms is currently being investigated \cite{sawatari}.  
In several fields, models have been constructed which include the
effects of time delays\cite{milton,biodelay,marcus,villermaux}. 
The method presented here
could possibly help in the critical examination of these models with extension
of including noise effects by 
comparisons with experimental time series, especially 
near the equilibrium state of the system. We are currently 
involved in the application of this and similar algorithms to
experimental time series from biological systems, such
as posture control data \cite{collins}, which can
physically be assumed to have a delayed feedback.

The authors would like to thank Prof. M. Tokoro of Keio University and
Sony CSL for providing an opportunity for this collaborative work.

\noindent

\begin{figure}
\dofig{2.0in}{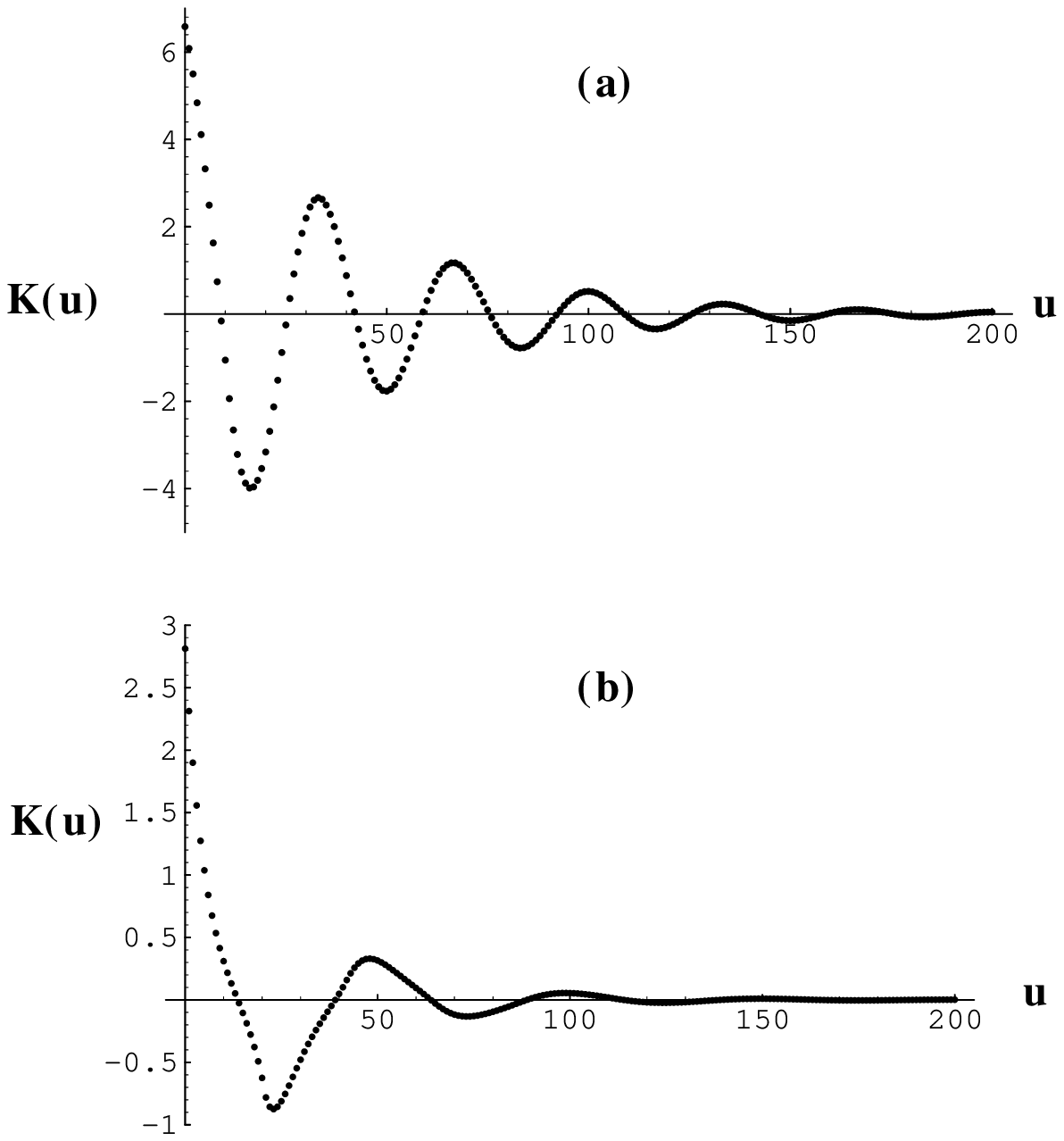}
\caption{Stationary correlation function $K(u)$  
as a function of steps $u$ iteratively obtained as
the solution of equation (3).
The parameters are set as 
$(\alpha, \beta, \tau) = (a) (0.1, 0.15, 10)$ and $(b) (0.2, 0.1, 20)$.}
\end{figure}

\begin{figure}
\dofig{2.0in}{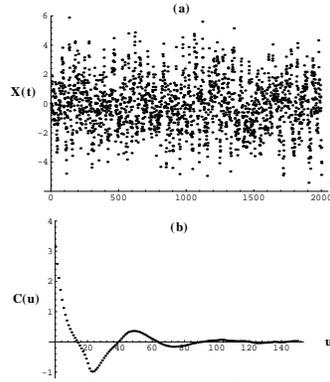}
\caption{
An example of time series (a) and associated correlation
function $C(u)$ generated from equation (4)(b).
The parameters are set as 
$(\alpha, \beta, \tau) =  (0.2, 0.1, 20)$.
}
\end{figure}

\begin{figure}
\dofig{2.0in}{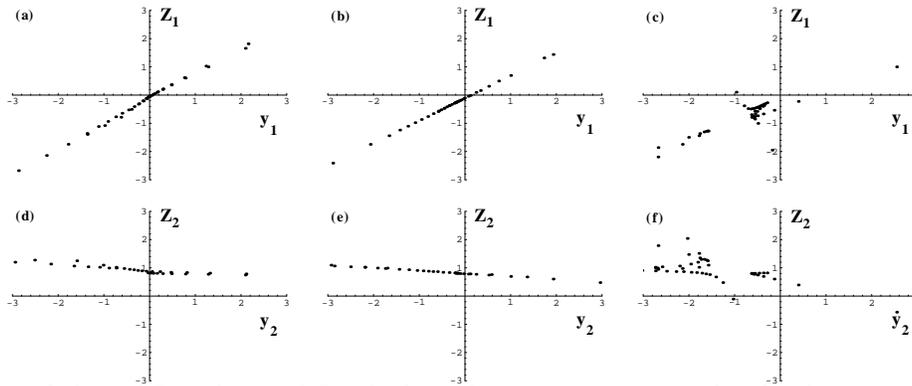}
\caption{
Examples of plots of $L_1$ (a,b,c) and $L_2$ (d,e,f),
with various estimation of $\tau_e$ 
for the time series shown in Fig. 2 with
$(\alpha, \beta, \tau) =  (0.2, 0.1, 20)$.
The estimates are 
$\tau_e = (a)(d) 15, (b)(e) 20, (c)(f) 25 $.
}
\end{figure}

\begin{figure}
\dofig{2.0in}{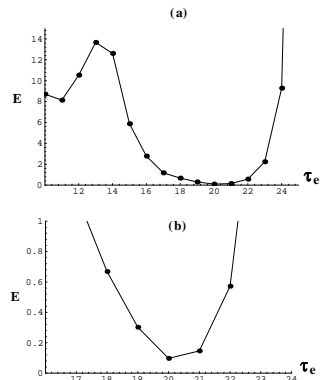}
\caption{
An example of plots of $E(\tau_e)$ with
various estimates around $\tau_i= 14$
for the time series shown in Fig. 2
with 
$(\alpha, \beta, \tau) =  (0.2, 0.1, 20)$.
(b) is with finer scale around the minimum of $E$.
}
\end{figure}
\clearpage

\begin{table}

\caption{Table of estimation results. For a time series with parameters of $(\alpha, \beta, \tau)$,
our method generates $\alpha_i, \beta_i$ and $\tau_e$, where $\alpha_i, \beta_i$ are estimated from
graph of $L_i, i=1,2$.}

\begin{tabular}{lll@{\hspace{0.7in}}lllllll}
$\alpha$ & $\beta$ & $\tau$  & $\tau_i$ & $\tau_e$ & $\alpha_1$ & $\alpha_2$
& $\beta_1$ & $\beta_2$ & $u_{max}$\\
\hline
  0.04 & 0.02 & 20 & 32 & 18 & 0.030 & 0.024 & 0.021 & 0.021 & 80\\
  0.05 & 0.02 & 50 & 41 & 50 & 0.051 & 0.051 & 0.020 & 0.022 & 80\\
  0.08 & 0.04 & 10 & 16 & 11 & 0.091 & 0.090 & 0.038 & 0.037 & 30\\
  0.10 & 0.15 &  5 &  6 &  5 & 0.10  & 0.10  & 0.15  & 0.15  & 20\\
  0.10 & 0.30 &  5 &  5 &  5 & 0.099 & 0.10  & 0.30  & 0.30  & 50\\
  0.20 & 0.10 & 20 & 14 & 20 & 0.21  & 0.20  & 0.10  & 0.098 & 50\\
  0.30 & 0.10 & 10 &  8 & 10 & 0.31  & 0.30  & 0.11  & 0.098 & 25\\
\end{tabular}
\end{table}

\end{document}